\documentclass[twocolumn,showpacs,preprintnumbers,amsmath,amssymb]{revtex4}
\usepackage{graphicx}% Include figure files
\usepackage{dcolumn}% Align table columns on decimal point
\usepackage{bm}% bold math

\begin{document}

\title{Parametrization of Neutrino Mixing Matrix in Tri-bimaximal Mixing Pattern}% Force line breaks with \\

\author{Nan Li}
\affiliation{School of Physics, Peking University, Beijing 100871,
China}%Lines break automatically or can be forced with \\
\author{Bo-Qiang Ma}\altaffiliation{Corresponding author}\email{mabq@phy.pku.edu.cn}
\affiliation{ CCAST (World Laboratory), P.O.~Box 8730, Beijing
100080, China \\
and School of Physics, Peking University, Beijing, 100871,
China\footnote{Mailing address}}

\begin{abstract}
The neutrino mixing matrix is expanded in powers of a small
parameter $\lambda$ in tri-bimaximal mixing pattern. We also
present some applications of this parametrization, such as to the
expression of the Jarlskog parameter $J$. Comparing with other
parametrizations (such as the parametrization in bimaximal mixing
pattern), this parametrization converges more quickly, but is of
less symmetry.
\end{abstract}

\pacs{14.60.Pq; 12.15.Ff; 14.60.Lm}

\maketitle

In recent years, the mixing of different generations of neutrinos
has been established by abundant experimental data. The
KamLAND~\cite{Kam} and SNO~\cite{sno} experiments showed that the
long-existed solar neutrino deficit is due to the oscillation from
$\nu_{e}$ to a mixture of $\nu_{\mu}$ and $\nu_{\tau}$ with a
mixing angle approximately of $\theta_{sol} \approx 34^{\circ}$.
The K2K~\cite{K2K} and Super-Kamiokande~\cite{SUPER} experiments
told us that the atmospheric neutrino anomaly is caused by the
$\nu_{\mu}$ to $\nu_{\tau}$ oscillation with almost the largest
mixing angle of $\theta_{atm} \approx 45^{\circ}$. On the other
hand, the non-observation of the disappearance of $\bar{\nu}_{e}$
in the CHOOZ~\cite{Chz} experiment indicated that the mixing angle
$\theta_{chz}$ is smaller than $5^{\circ}$ at the best fit
point~\cite{garcia,Altarelli}.

These experiments not only confirmed the oscillations of
neutrinos, but also measured the mass-squared differences of the
neutrino mass eigenstates~\cite{garcia} (the allowed ranges at
3$\sigma$), $1.6\times10^{-3}~\mbox{eV}^{2}<\Delta
m_{atm}^{2}=|m_{3}^{2}-m_{2}^{2}|<3.6\times
10^{-3}~\mbox{eV}^{2}$, and $7.3\times10^{-5}~\mbox{eV}^{2}<\Delta
m_{sol}^{2}=|m_{2}^{2}-m_{1}^{2}|<9.3\times
10^{-5}~\mbox{eV}^{2}$, where $\pm$ correspond to the normal and
inverted mass schemes respectively.

Just like the Cabibbo-Kobayashi-Maskawa (CKM)~\cite{ckm} matrix
for quark mixing, the neutrino mixing matrix is described by the
unitary Pontecorvo-Maki-Nakawaga-Sakata (PMNS)~\cite{mns} matrix
$V$, which links the neutrino flavor eigenstates $\nu_{e}$,
$\nu_{\mu}$, $\nu_{\tau}$ to the mass eigenstates $\nu_{1}$,
$\nu_{2}$, $\nu_{3}$,
\begin{eqnarray}%\nonumber
 \left(
    \begin {array}{c}
       \nu_{e} \\
       \nu_{\mu} \\
       \nu_{\tau} \\
\end{array}
\right) = \left(
    \begin {array}{ccc}
       V_{e1} & V_{e2} & V_{e3}\\
       V_{\mu1} & V_{\mu2} & V_{\mu3}\\
       V_{\tau1} & V_{\tau2} & V_{\tau3}\\
\end{array}
\right) \left(
    \begin {array}{c}
       \nu_{1} \\
       \nu_{2} \\
       \nu_{3} \\
\end{array}
\right). \nonumber
\end{eqnarray}

If neutrinos are of Dirac type, the neutrino mixing matrix can be
written as follows (with three mixing angles and a Dirac
CP-violating phase, analogous to that of quarks)
\begin{eqnarray}%\nonumber
    V=\left(
        \begin{array}{ccc}
            c_{2}c_{3} & c_{2}s_{3} & s_{2}e^{-i\delta}\\
           -c_{1}s_{3}-s_{1}s_{2}c_{3}e^{i\delta} & c_{1}c_{3}-s_{1}s_{2}s_{3}e^{i\delta} & s_{1}c_{2}\\
            s_{1}s_{3}-c_{1}s_{2}c_{3}e^{i\delta} & -s_{1}c_{3}-c_{1}s_{2}s_{3}e^{i\delta} & c_{1}c_{2}\\
        \end{array}
        \right),
        \nonumber
\end{eqnarray}
where $s_{i}=\sin\theta_{i}$, $c_{i}=\cos\theta_{i}$ (for $i=1, 2,
3$), and $\delta$ is the Dirac CP-violating phase. If neutrinos
are of Majorana type, it is always feasible to parametrize the
neutrino mixing matrix as a product of the Dirac neutrino mixing
matrix and a diagonal phase matrix with two unremovable phase
angles $\mbox{diag} (e^{i\alpha}, e^{i\beta}, 1)$~\cite{sch},
where $\alpha$, $\beta$ are the Majorana CP-violating phases. The
Dirac CP-violating phase is associated with the neutrino
oscillations, CP and T violation, and the Majorana CP-violating
phases are associated with the neutrinoless double beta decay, and
lepton-number-violating processes~\cite{sch1}.

The three mixing angles $\theta_{atm}$, $\theta_{chz}$, and
$\theta_{sol}$ are related to $\theta_{1}$, $\theta_{2}$, and
$\theta_{3}$, which describe the mixing between 2nd and 3rd, 3rd
and 1st, 1st and 2nd generations of neutrinos. To a good degree of
accuracy, $\theta_{atm}=\theta_{1}$, $\theta_{chz}=\theta_{2}$,
and $\theta_{sol}=\theta_{3}$.

According to the results of the global analysis of the neutrino
oscillation experimental data, the elements of the modulus of the
neutrino mixing matrix are summarized as follows~\cite{garcia}
\begin{equation}
    |V|=\left(
        \begin{array}{ccc}
             0.77-0.88 & 0.47-0.61 & <0.20\\
             0.19-0.52 & 0.42-0.73 & 0.58-0.82\\
             0.20-0.53 & 0.44-0.74 & 0.56-0.81\\
        \end{array}
        \right),
\end{equation}
and the best fit points of the modulus of $V$ are~\cite{Altarelli}
\begin{equation}
    |V|=\left(
        \begin{array}{ccc}
             0.84 & 0.54 & 0.08\\
             0.44 & 0.56 & 0.72\\
             0.32 & 0.63 & 0.69\\
        \end{array}
        \right).
\end{equation}

Quite different from quark mixing matrix, almost all the
non-diagonal elements of the neutrino mixing matrix are large,
only with the exception of  $V_{e3}$. So it is unpractical to
expand the matrix in powers of one of the non-diagonal elements,
like the Wolfenstein parametrization~\cite{wol} of the quark
mixing matrix. The quark mixing matrix is very near the unit
matrix, but it is not a small modification to the unit matrix in
neutrino mixing pattern. Several bases of neutrino mixing matrix
are summarized as follows~\cite{xingzz} (they all take some of the
mixing angles as special values.)
%\begin{widetext}
\begin{eqnarray}
    \left(
        \begin{array}{ccc}
            \sqrt{2}/2 & \sqrt{2}/2 & 0 \\
            -\sqrt{6}/6 & \sqrt{6}/6 & \sqrt{6}/3 \\
            \sqrt{3}/3 & -\sqrt{3}/3 & \sqrt{3}/3
        \end{array}
        \right),
    \left(
        \begin{array}{ccc}
            \sqrt{3}/2 & 1/2 & 0 \\
            -\sqrt{2}/4 & \sqrt{6}/4 & \sqrt{2}/2 \\
            \sqrt{2}/4 & -\sqrt{6}/4 & \sqrt{2}/2
        \end{array}
        \right), \nonumber \\
    \left(
        \begin{array}{ccc}
            \sqrt{2}/2 & \sqrt{2}/2 & 0 \\
            -1/2 & 1/2 & \sqrt{2}/2 \\
            1/2 & -1/2 & \sqrt{2}/2
        \end{array}
        \right),
    \left(
        \begin{array}{ccc}
            \sqrt{6}/3 & \sqrt{3}/3 & 0 \\
            -\sqrt{6}/6 & \sqrt{3}/3 & \sqrt{2}/2 \\
            \sqrt{6}/6 & -\sqrt{3}/3 & \sqrt{2}/2
        \end{array}
        \right).\nonumber
\end{eqnarray}
%\end{widetext}
Therefore we may expand the neutrino mixing matrix around these
bases. The third matrix is the bimaximal mixing pattern, and the
expansions around it have been discussed by
Rodejohann~\cite{Rodejohann}, Giunti and Tanimoto~\cite{Giunti},
and us~\cite{li}.

The fourth matrix is the tri-bimaximal pattern. It was first
conjectured by Wolfenstein~\cite{Wolfenstein1978} long ago, and
was discussed by several other authors recently~\cite{he}. It is
the best approximation to the neutrino mixing matrix, and its
three mixing angles are $45^{\circ}$, $0^{\circ}$ and
$35.3^{\circ}$, which agree with the experimental data perfectly.
So in this paper, we will expand the neutrino mixing matrix around
it.

Comparing with Eq.~(2), we can make an expansion of $V$ in powers
of $\lambda$, which satisfies
\begin{equation}
  V_{e2}={\sqrt{3}}/{3}-\lambda,
\end{equation}
where $\lambda$ measures the strength of the deviation of $V_{e2}$
from the tri-bimaximal mixing pattern. Because the best fit point
of $V_{e2}$ is 0.53~\cite{garcia}, $\lambda$ is a small parameter,
which approximately equals to 0.05, and this expansion is
reasonable and will converge quickly.

For the range of $\lambda$, from the analyses of the experimental
data~\cite{garcia}, we have $0.51<V_{e2}<0.55$ (the allowed range
at 1$\sigma$). So $0.51<\sqrt{3}/3-\lambda<0.55$, and we can get
$0.03<\lambda<0.07$. Similarly, $-0.03<\lambda<0.1$ (the allowed
range at 3$\sigma$).

Similarly, with the global analyses on the experimental data, the
best fit point of $|V_{\mu3}|^2$ is 0.52~\cite{Altarelli}.
Therefore we have $V_{\mu3}=0.72$, and the deviation of $V_{\mu3}$
from $\sqrt{2}/2$ is very small, so we can set
\begin{equation}
V_{\mu3}={\sqrt{2}}/{2}+a\lambda.
\end{equation}
Thus $a\lambda\sim0.013$, and $a\sim0.3$.

Furthermore, since $\theta_{2}$ is rather small (with the global
analyses, $|V_{e3}|<0.25$ (the allowed range at 3$\sigma$), and
with the best fit point $|V_{e3}|=0.08$
~\cite{Altarelli},~\cite{Altarelli2}), we can set
\begin{equation}
V_{e3}=b\lambda e^{i\delta}.
\end{equation}
So $b\sim1.5$. Due to the uncertainty of the value of
$\theta_{2}$, only the upper bound 0.25 is meaningful in
phenomenological anlyses, and the parametrization in Eq.~(5) is
only an assumption, however, we can adjust the value of $b$ to
satisfy the best fit point of $V_{e3}$, which can be determined by
the long baseline experiments~\cite{long} in the future.

%Here we can see that both $a$ and $b$ are small parameters of order 1.

Altogether, there are four parameters here, $a$, $b$, $\lambda$
and $\delta$. They can describe the neutrino mixing matrix
completely, both the real and the imaginary parts.

Now we will calculate all the $s_{i}$ and $c_{i}$ (for $i=1, 2,
3$) to the order of $\lambda^{3}$. From Eq.~(5), $s_{2}=b\lambda$,
we have
\begin{equation}
c_{2}=\sqrt{1-s_{2}^{2}}=1-\frac{1}{2}b^{2}\lambda^{2}.
\end{equation}
 From Eq.~(4), we have
\begin{eqnarray}
%\nonumber
s_{1}c_{2}=V_{\mu3}={\sqrt{2}}/{2}+a\lambda, \nonumber
\end{eqnarray}
using Eq.~(6), we get
\begin{equation}
s_{1}=\frac{\sqrt{2}}{2}+a\lambda+\frac{\sqrt{2}}{4}b^{2}\lambda^{2}+\frac{1}{2}ab^2\lambda^{3}.
\end{equation}
Similarly,
\begin{eqnarray}
c_{1}&=&\frac{\sqrt{2}}{2}-a\lambda-(\sqrt{2}a^{2}+\frac{\sqrt{2}}{4}b^{2})\lambda^{2}-(2a^3+\frac{3}{2}ab^2)\lambda^{3},\nonumber\\
s_{3}&=&\frac{\sqrt{3}}{3}-\lambda+\frac{\sqrt{3}}{6}b^2\lambda^{2}-\frac{1}{2}b^2\lambda^{3},\nonumber\\
c_{3}&=&\frac{\sqrt{6}}{3}+\frac{\sqrt{2}}{2}\lambda-(\frac{3\sqrt{6}}{8}+\frac{\sqrt{6}}{12}b^{2})\lambda^{2}
\nonumber\\&&
+(\frac{9\sqrt{2}}{16}+\frac{5\sqrt{2}}{8}b^{2})\lambda^{3}.
\end{eqnarray}
Thus we obtain all the trigonometric functions of the three mixing
angles.

So we can get all the elements of the neutrino mixing matrix
straightforwardly (to the order of $\lambda^{2}$),
\begin{eqnarray}%\nonumber
V_{e1}&=&\frac{\sqrt{6}}{3}+\frac{\sqrt{2}}{2}\lambda-(\frac{3\sqrt{6}}{8}+\frac{\sqrt{6}b^2}{4})\lambda^2,\nonumber\\
V_{e2}&=&\frac{\sqrt{3}}{3}-\lambda, %\\%
\nonumber\\
V_{e3}&=&b\lambda e^{i\delta}, \\ %\nonumber \\
%\end{eqnarray}
%\begin{eqnarray}
V_{\mu1}&=&-\frac{\sqrt{6}}{6}+(\frac{\sqrt{2}}{2}+\frac{\sqrt{3}a}{3})\lambda-(a-\frac{\sqrt{6}a^2}{3})\lambda^{2}\nonumber \\
&&-[\frac{\sqrt{3}}{3}b\lambda+(\frac{b}{2}+\frac{\sqrt{6}ab}{3})\lambda^{2}] e^{i\delta} , %\\%
\nonumber\\
V_{\mu2}&=&\frac{\sqrt{3}}{3}+(\frac{1}{2}-\frac{\sqrt{6}a}{3})\lambda-(\frac{3\sqrt{3}}{8}+\frac{\sqrt{2}a}{2}+\frac{2\sqrt{3}a^2}{3}\nonumber\\
&&+\frac{\sqrt{3}b^2}{4})\lambda^{2}-[\frac{\sqrt{6}}{6}b\lambda-(\frac{\sqrt{2}b}{2}-\frac{\sqrt{3}ab}{3})\lambda^2] e^{i\delta}, \nonumber \\
V_{\mu3}&=&\frac{\sqrt{2}}{2}+a\lambda, \nonumber \\
%\end{eqnarray}
%\begin{eqnarray}
V_{\tau1}&=&\frac{\sqrt{6}}{6}-(\frac{\sqrt{2}}{2}-\frac{\sqrt{3}a}{3})\lambda-(a-\frac{\sqrt{6}b^2}{6})\lambda^{2}\nonumber \\
&&-[\frac{\sqrt{3}}{3}b\lambda+(\frac{b}{2}-\frac{\sqrt{6}ab}{3})\lambda^2] e^{i\delta},\nonumber\\
V_{\tau2}&=&-\frac{\sqrt{3}}{3}-(\frac{1}{2}+\frac{\sqrt{6}a}{3})\lambda+(\frac{3\sqrt{3}}{8}-\frac{\sqrt{2}a}{2}-\frac{\sqrt{3}b^2}{12})\lambda^{2}\nonumber \\
&&-[\frac{\sqrt{6}}{6}b\lambda-(\frac{\sqrt{2}b}{2}+\frac{\sqrt{3}ab}{3})\lambda^2] e^{i\delta}, %\\%
\nonumber \\
V_{\tau3}&=&\frac{\sqrt{2}}{2}-a\lambda-(\sqrt{2}a^{2}+\frac{\sqrt{2}}{2}b^{2})\lambda^{2}.
\nonumber
\end{eqnarray}

Then we can expand the neutrino mixing matrix in powers of
$\lambda$ (to the order of $\lambda^2$),
\begin{widetext}
\begin{eqnarray}\nonumber
   V&=&\left(
        \begin{array}{ccc}
            \frac{\sqrt{6}}{3} & \frac{\sqrt{3}}{3} & 0 \\
            -\frac{\sqrt{6}}{6} & \frac{\sqrt{3}}{3} & \frac{\sqrt{2}}{2} \\
            \frac{\sqrt{6}}{6} & -\frac{\sqrt{3}}{3} & \frac{\sqrt{2}}{2}
        \end{array} \right)+\lambda
 \left(
        \begin{array}{ccc}
            \frac{\sqrt{2}}{2} & -1 & b e^{i\delta} \\
            (\frac{\sqrt{2}}{2}+\frac{\sqrt{3}a}{3})-\frac{\sqrt{3}}{3}b e^{i\delta} & (\frac{1}{2}-\frac{\sqrt{6}a}{3})-\frac{\sqrt{6}}{6}b e^{i\delta} & a \\
            -(\frac{\sqrt{2}}{2}-\frac{\sqrt{3}a}{3})-\frac{\sqrt{3}}{3}b e^{i\delta} & -(\frac{1}{2}+\frac{\sqrt{6}a}{3})-\frac{\sqrt{6}}{6}b e^{i\delta} & -a
       \end{array} \right)\\\nonumber&&+\lambda^{2}
 \left(
        \begin{array}{ccc}
            -(\frac{3\sqrt{6}}{8}+\frac{\sqrt{6}b^2}{4}) & 0 & 0 \\
            -(a-\frac{\sqrt{6}a^2}{3})-(\frac{b}{2}+\frac{\sqrt{6}ab}{3}) e^{i\delta} & -(\frac{3\sqrt{3}}{8}+\frac{\sqrt{2}a}{2}+\frac{2\sqrt{3}a^2}{3}+\frac{\sqrt{3}b^2}{4})+(\frac{\sqrt{2}b}{2}-\frac{\sqrt{3}ab}{3}) e^{i\delta} & 0 \\
            -(a-\frac{\sqrt{6}b^2}{6})-(\frac{b}{2}-\frac{\sqrt{6}ab}{3}) e^{i\delta} &  (\frac{3\sqrt{3}}{8}-\frac{\sqrt{2}a}{2}-\frac{\sqrt{3}b^2}{12})+(\frac{\sqrt{2}b}{2}+\frac{\sqrt{3}ab}{3}) e^{i\delta} & -(\sqrt{2}a^{2}+\frac{\sqrt{2}}{2}b^{2})
        \end{array} \right)\\&&+\cdots. \nonumber
\end{eqnarray}
%\nonumber
\end{widetext}

Now we will see the meaning of every order in the expansion of
$V$.

1. The term of $\lambda^{0}$ is the approximation of the lowest
order, where the mixing angles are of $45^{\circ}$, $0^{\circ}$
and $35.3^{\circ}$. We call this the tri-bimaximal mixing pattern,
and it is nearest to the experimental data among the bases with
special mixing angles.

2. The term of $\lambda^{1}$ indicates the deviation of the
neutrino mixing matrix from the tri-bimaximal mixing pattern. Also
it shows the effect of CP violation. Because CP violation is
described by the element $V_{e3}$~\cite{prdd}, and in the terms of
$\lambda^{0}$, $V_{e3}=0$, the degree of CP violation is of the
order $\lambda^{1}$ in our parametrization.

3. The term of $\lambda^{2}$ and so on are the modifications of
higher orders.

In this parametrization, several other corresponding observable
quantities associated with the elements of the neutrino mixing
matrix can be expressed in relatively simple forms.

1. The Jarlskog parameter $J$~\cite{Ja}. $J$ is the
rephasing-invariant measurement of the lepton CP violation. The
Majorana  CP-violating phases can be removed away by redefining
the phases of the Dirac fields, so only $\delta$ is associated
with CP violation.
$J=\mbox{Im}(V_{e2}V_{\mu3}V_{e3}^{\ast}V_{\mu2}^{\ast})=s_{1}s_{2}s_{3}c_{1}c_{2}^2c_{3}\sin\delta$.
In our parametrization, $J$ can be expressed in a simple form (to
the order of $\lambda^{2}$),
\begin{equation}
J=\frac{\sqrt{2}}{6}b\lambda\sin\delta(1-\frac{\sqrt{3}}{2}\lambda).
\end{equation}
Because $s_{1}$, $c_{1}$, $s_{3}$, $c_{3}$ have the factors
$\frac{\sqrt{2}}{2}$, $\frac{\sqrt{2}}{2}$, $\frac{\sqrt{3}}{3}$,
$\frac{\sqrt{6}}{3}$, there are four factors smaller than 1 in
$J$. So the degree of the lepton CP violation is suppressed four
times, $(\frac{\sqrt{2}}{2})^2 \frac{\sqrt{3}}{3}
\frac{\sqrt{6}}{3}=\frac{\sqrt{2}}{6}$. Again, $J$ is suppressed
by the factor $b\lambda\sim0.08$~\cite{Altarelli}. We can
determine the range of $J$, $J\sim 0.018$. (here we take
$\lambda\sim0.05$ and $\sin\delta\sim1$.)

2. The effective Majorana mass term $\langle m \rangle_{ee}$. In
the neutrinoless double beta decay, the effective Majorana mass
term is defined as follows
\begin{eqnarray}\nonumber
\langle
m\rangle_{ee}\equiv|m_{1}V_{e1}^{2}e^{2i\alpha}+m_{2}V_{e2}^{2}e^{2i\beta}+m_{3}V_{e3}^{2}|.
\end{eqnarray}
Using Eq.~(9), we get
\begin{eqnarray}\nonumber
\langle m
\rangle_{ee}&=&|\frac{1}{3}(2m_{1}e^{2i\alpha}+m_{2}e^{2i\beta})\nonumber
\\&&+\frac{2\sqrt{3}}{3}\lambda(m_{1}e^{2i\alpha}-m_{2}e^{2i\beta}) \nonumber \\
&&-\lambda^{2}[(m_{1}e^{2i\alpha}-m_{2}e^{2i\beta})+b^2(m_{1}e^{2i\alpha}-m_{3}e^{2i\delta})]|.\nonumber
\end{eqnarray}
We can see that the coefficients of the three terms show the
influences of the three orders of $\lambda$. Only $m_{1}$ and
$m_{2}$ are important to the value of $\langle m \rangle_{ee}$,
and the influence of $m_{3}$ almost vanish if the masses of the
three mass eigenstates are nearly degenerated, because the
coefficient of it $b^{2}\lambda^{2}$ is of $10^{-3}$.

3. The effective mass terms of neutrinos. The effective mass terms
of neutrinos can be defined as follows (here we take electron
neutrino for example.)
\begin{eqnarray}\nonumber
\langle m\rangle_{e}^{2}\equiv
m_{1}^{2}|V_{e1}|^{2}+m_{2}^{2}|V_{e2}|^{2}+m_{3}^{2}|V_{e3}|^{2}.
\end{eqnarray}
Using Eq.~(9), we get
\begin{eqnarray}
\langle m\rangle_{e}^{2}
&=&\frac{1}{3}(2m_{1}^{2}+m_{2}^{2})-\frac{2\sqrt{3}}{3}\lambda(m_{2}^{2}-m_{1}^{2})\nonumber\\
&&+\lambda^{2}[(m_{2}^{2}-m_{1}^{2})+b^2(m_{3}^{2}-m_{1}^{2})].
\end{eqnarray}
Again, the coefficients of the three terms show the influences of
the three orders of $\lambda$. Noting that $\Delta
m_{sol}^{2}=|m_{2}^{2}-m_{1}^{2}|$ and $\Delta
m_{atm}^{2}=|m_{3}^{2}-m_{2}^{2}|$, we can rewrite Eq.~(11) into
\begin{eqnarray}
\langle m \rangle_{e}^{2}&=&
m_{1}^{2}+[\frac{1}{3}-\frac{2\sqrt{3}}{3}\lambda+(b^2+1)\lambda^2](m_{2}^{2}-m_{1}^{2})\nonumber\\
&&+b^2\lambda^2(m_{3}^{2}-m_{2}^{2})\nonumber \\
&=&m_{1}^{2}\pm
[\frac{1}{3}-\frac{2\sqrt{3}}{3}\lambda+(b^2+1)\lambda^2]\Delta
m_{sol}^{2}\nonumber\\
&&\pm b^2\lambda^2\Delta m_{atm}^{2},
\end{eqnarray}
where the first sign of ``$\pm$" should be chosen as ``+" if we
accept $m_2>m_1$ because of Mikheyev-Smirnov-Wolfenstein (MSW)
~\cite{msw} matter effect on solar neutrinos. We can see from
Eq.~(12) that $\langle m \rangle_{e}^{2}$ is directly related with
the masses and the mass-squared differences of neutrinos. So these
two kinds of different observable quantities are associated
together in our parametrization. If we can separately measure
$\Delta m_{atm}^{2}$, $\Delta m_{sol}^{2}$, and $\langle m
\rangle_{e}^{2}$ to a good degree of accuracy, we can fix the
value of $m_{1}$, which will help us determine the absolute mass
of neutrino ultimately.

Finally, we will give some discussion and comparison between the
different methods in parametrizing the neutrino mixing matrix.

For the quark mixing, all the non-diagonal elements are small, so
it is practical to expand the quark mixing matrix around the unit
matrix. But the case is clearly different for the neutrino mixing.
So if we still present Wolfenstein-like parametrization for the
neutrino mixing matrix (as Xing did~\cite{xingzhizhong}), we have
to use much higher orders of the non-diagonal elements. Hence it
is necessary for us to find new bases for the expansion of the
neutrino mixing matrix.

Among all the matrices with special mixing angles, the
tri-biamaximal mixing pattern seems to be the best one. So it is
natural to expand the neutrino mixing matrix around it. This is
the main point of our paper. But just due to the smallness of
$\lambda$ ($\lambda\sim0.05$), the expansion converges so quickly
that the modulus of the matrix is rather small only to the order
of $\lambda^2$. However, in the bimaximal case, we can expand the
neutrino mixing matrix to higher orders, and can see different
physical effects in different orders, because $\lambda$ there is
larger ($\lambda\sim0.1$)~\cite{li}. Moreover, the expansion in
tri-bimaximal mixing pattern has less symmetry than the expansion
in bimaximal mixing pattern, because the mixing angles are not the
same here.

Altogether, we can see that there are advantages and deficiencies
at the same time in both the expansions in tri-bimaximal and
bimaxiamal mixing patterns, and the adoption of which of them
should be determined by more and more precise experimental data.

In summary, although all sorts of parametrization of the neutrino
mixing matrix are not based on any deep theoretical foundation and
are equivalent mathematically, and applying any of them is
arbitrary, however, it is quite likely that some particular
parametrization is useful in making sense of experimental data.
Furthermore, we can express some other observable quantities in a
relatively simple form, and can link several different kinds of
observable quantities together. This is the purpose of our
parametrization.

\begin{acknowledgments}
We are very grateful to Prof. Xiao-Gang He for his stimulating
suggestions and discussions. This work is partially supported by
National Natural Science Foundation of China under Grant Numbers
10025523 and 90103007.
\end{acknowledgments}

%\newpage


\begin{thebibliography}{99}
\bibitem{Kam}
KamLAND Collaboration, K.~Eguchi {\it et al.}, Phys. Rev. Lett.
{\bf 90}, 021802 (2003).

\bibitem{sno}
SNO Collaboration, S.N.~Ahmed {\it et al.}, Phys. Rev. Lett. {\bf
92}, 181301 (2004).

\bibitem {K2K}
K2K Collaboration, M.H.~Ahn {\it et al.}, Phys. Rev. Lett. {\bf
90}, 041801 (2003).

\bibitem {SUPER}
C.K.~Jung, C.~McGrew, T.~Kajita, and T.~Mann, Anna. Rev. Nucl.
Part. Sci. {\bf 51}, 451 (2001); Super-Kamiokande Collaboration,
Y.~Hayato, talk at HEP2003 International Europhysics Conference,
Aachen, Germany, 2003, http://eps2003.physik.rwth-aachen.de

\bibitem{Chz}
CHOOZ Collaboration, M.~Apollonio {\it et al.}, Phys. Lett. {\bf B
420}, 397 (1998); Palo Verde Collaboration, F.~Boehm {\it et al.},
Phys. Rev. Lett. {\bf 84}, 3764 (2000).

\bibitem{garcia}
M.C.~Gonzalez-Garcia, hep-ph/0410030.

\bibitem{Altarelli}
G.~Altarelli, hep-ph/0410101.

\bibitem{Altarelli2}
G.~Altarelli and F.~Feruglio, hep-ph/0405048, New J. Phys. {\bf
6}, 106 (2004).

\bibitem{ckm}
N.~Cabibbo, Phys. Rev. Lett. {\bf 10}, 531 (1963); M.~Kobayashi
and T.~Maskawa, Prog. Theor. Phys. {\bf 149}, 652 (1973).

\bibitem{mns}
B.~Pontecorvo, Sov. Phys. JETP {\bf 7}, 172 (1958); Z.~Maki,
M.~Nakawaga, and S.~Sakata, Prog. Theor. Phys. {\bf 28}, 870
(1962).

\bibitem{sch}
J.~Schechter and J.W.F.~Valle, Phys. Rev. {\bf D 22}, 2227 (1980);
S.M.~Bilenky, J.~Hosek, and S.T.~Petsov, Phys. Lett. {\bf B 94},
495 (1980).

\bibitem{sch1}
J.~Schechter and J.W.F.~Valle, Phys. Rev. {\bf D 23}, 1666 (1981);
Phys. Rev. {\bf D 24}, 1883 (1981); Phys. Rev. {\bf D 25}, 283
(1982).

\bibitem{wol}
L.~Wolfenstein, Phys. Rev. Lett. {\bf 51}, 1945 (1983).

\bibitem{xingzz}
Z.Z.~Xing, Int. J. Mod. Phys. {\bf A 19}, 1 (2004).

\bibitem{Rodejohann}
W.~Rodejohann, Phys. Rev. {\bf D 69}, 033005 (2004); Nucl. Phys.
{\bf B 687}, 31 (2004).

\bibitem{Giunti}
C.~Giunti and M.~Tanimoto, Phys. Rev. {\bf D 66}, 053013 (2002);
Phys. Rev. {\bf D 66}, 113006 (2002).

\bibitem{li}
N.~Li and B.-Q.~Ma, Phys. Lett. {\bf B 600}, 248 (2004).

\bibitem{Wolfenstein1978}
L.~Wolfenstein, Phys. Rev. {\bf D 18}, 958 (1978).

\bibitem{he}
P.F.~Harrison, D.H.~Perkins, and W.G.~Scott, Phys. Lett. {\bf B
530}, 167 (2002); Z.Z.~Xing, Phys. Lett. {\bf B 533}, 85 (2002);
X.G.~He and A.~Zee, Phys. Lett. {\bf B 560}, 87 (2003); C.I.~Low
and R.R.~Volkas, Phys. Rev. {\bf D 68}, 033007 (2003); A.~Zee,
Phys. Rev. {\bf D 68}, 093002 (2003).

\bibitem{long}
%H.~Minakata {\it et al.},
H.~Minakata,H.~Sugiyama, O.~Yasuda, K.~Inoue, and F.~Suekane,
Phys. Rev. {\bf D 68} 033017 (2003); P.~Huber {\it et al.}, Nucl.
Phys. {\bf B 665} 487 (2003).

\bibitem{prdd}
F.J.~Gilman, K.~Kleinknecht, and B.~Renk, Phys. Rev. {\bf D 66},
010001 (2002); L.~Wolfenstein, Phys. Rev. {\bf D 66}, 010001
(2002).

\bibitem{Ja}
C.~Jarlskog, Phys. Rev. Lett. {\bf 55}, 1039 (1985); Z. Phys. {\bf
C 29}, 491 (1985).

\bibitem{msw}
S.P.~Mikheyev and A.Yu.~Smirnov, Sov. J. Nucl. Phys. {\bf 42} 913
(1985); L.~Wolfenstein, Phys. Rev. {\bf D 17} 2369 (1978).

\bibitem{xingzhizhong}
Z.Z.~Xing, J. Phys. {\bf G 29}, 2227 (2003).


\end{thebibliography}
\end{document}